%% file: paper.tex
\documentclass[10pt,twocolumn,letterpaper]{article}

\usepackage{cvpr}
\usepackage{times}
\usepackage{epsfig}
\usepackage{graphicx}
\usepackage{amsmath}
\usepackage{amssymb}
\usepackage{epigraph}
\usepackage{authblk}

\usepackage[pagebackref=true,breaklinks=true,letterpaper=true,colorlinks,bookmarks=false]{hyperref}

 \cvprfinalcopy 

\def\cvprPaperID{298} 

\ifcvprfinal\fi

\begin{document}

\title{Audio to Body Dynamics}


\author[1,3]{Eli Shlizerman}
\author[1,2]{Lucio Dery}
\author[1]{Hayden Schoen}
\author[1,3]{Ira Kemelmacher-Shlizerman}
\affil[1]{Facebook Inc.}
\affil[2]{Stanford University}
\affil[3]{University of Washington}

\teaser{
    	\includegraphics[width=.99\textwidth]{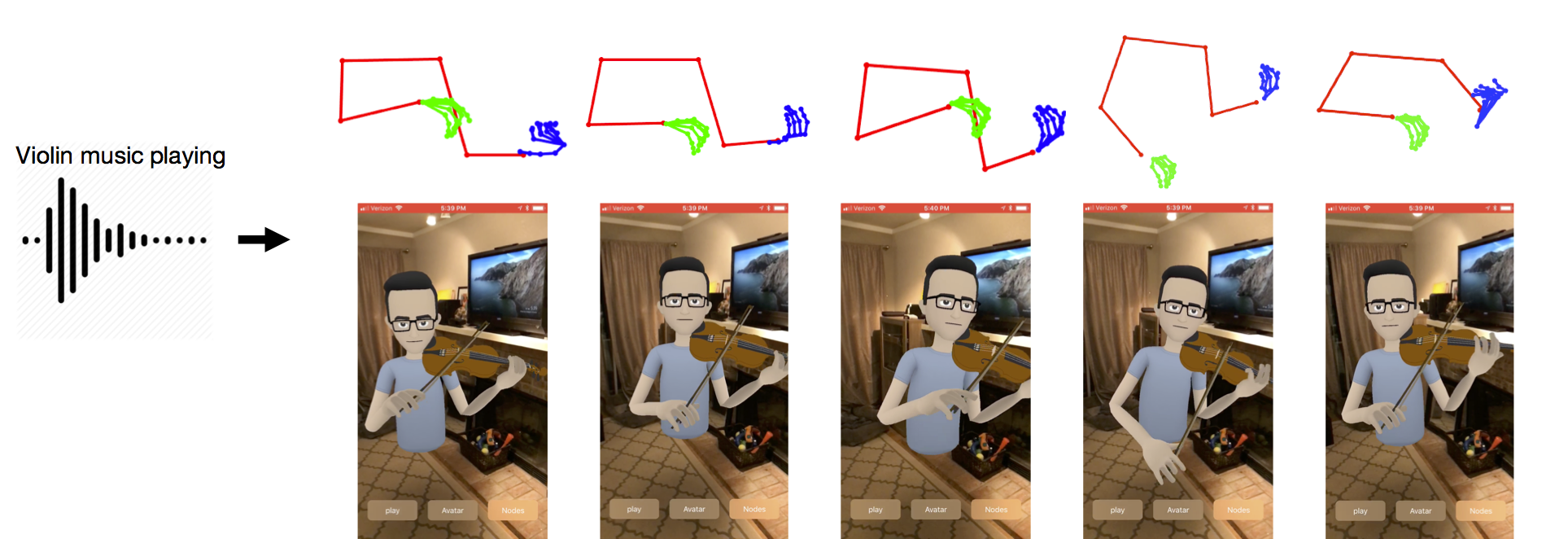} \caption{Given as input a music of violin (audio signal), our method (1) predicts body skeleton and (2) uses the skeleton to animate an avatar. See videos in the supplementary material (with audio on). } \label{fig:teaser} 
    	}

\maketitle

\begin{abstract}
We present a method that gets as input an audio of violin or piano playing, and outputs a video of skeleton predictions which are further used to animate an avatar. The key idea is to create an animation  of an avatar that moves their hands similarly to how a pianist or violinist would do,  just from audio.   Aiming for a fully detailed correct arms and fingers motion is a goal, however, it's not  clear if body movement can be predicted from music at all.  In this paper, we present the first result that shows that natural body dynamics can be predicted at all.  We built  an LSTM network that is trained  on violin and piano recital videos uploaded to the Internet. The predicted points are applied onto a rigged avatar to create the animation. 

\end{abstract}

\input{content/intro.tex}

\input{content/related.tex}
\input{content/data.tex}

\input{content/neuralnet.tex}

\input{content/avatar.tex}

\input{content/results.tex}

\input{content/conclusion.tex}

{\small
\bibliographystyle{ieee}
\bibliography{paper}
}

\end{document}

%% file: content/intro.tex
\section{Introduction}



\epigraph{All the same it is being said everywhere that I played too softly, or rather, too delicately for people used to the piano-pounding of the artists here.}{\textit{Frederic Chopin}}

When pianists play a musical piece on a piano, their body  reacts to the music. Their fingers  strike  piano keys to create  music. They move their arms to play on different octaves.  Violin players draw the bow with one hand across the strings and touch lightly or pluck the strings with the other hand's fingers.  Faster bowing produces faster music pace. 

An interesting question is: can body movement be  predicted computationally from a music signal?  This is a highly challenging computational problem. We need to have a good training set of videos, we need to be able to accurately predict body poses in those videos, and build an algorithm that is able to find correlation between music and body, to further  predict movement.

\begin{figure*}
\includegraphics[width=.99\textwidth]{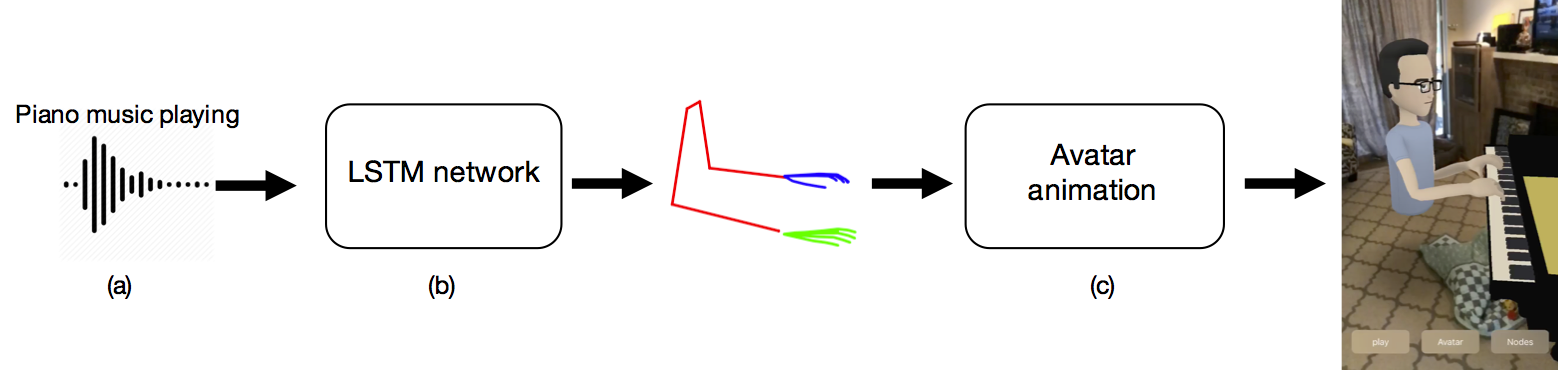}
\caption{Method overview: (a) Our method gets as input an audio signal, e.g., piano music,  (b) that is fed into our LSTM network to predict body movement points, (c) which in turn are used to animate an avatar and show it playing the input music on a piano (the avatar and piano are models while the rest is real apartment background). }
\label{fig:method_overview}
\end{figure*}

Human body dynamics is complex, particularly given the quality needed to learn correlation to audio. Traditionally, state of the art prediction of natural body movement from video sequences (not audio) used laboratory captured motion capture sequences. E.g., in our scenario we would need to bring a pianist to a laboratory and have them play several hours with sensors attaches to their fingers and body joints. Such approach is hard to execute in practice and not easily generalizable. If we could leverage the publicly available videos of highly skilled people playing online we potentially allow higher degree of diversity in data. Until recently, though estimating accurate body pose from videos was not possible. This year, several methods appeared that may allow us to learn from data ``in the wild''. 

In parallel, a number of methods \cite{suwajanakorn2017synthesizing,taylor2017deep,karras2017audio} also showed remarkable results for lip sync from speech. I.e., given an audio of a person saying a sentence they showed that it is possible to predict how that person's mouth landmarks would move while saying the words.  

These two advancements inspired us to tackle the challenging ideas of predicting body and fingers movement just from music. The goal of this paper is to explore if it's possible at all, and if we can create natural and logical \textbf{body} dynamics from audio.  Note that we do not use information like midi files from which we can potentially learn the correlation between exact piano keys and music.  We focus on creating an avatar that moves their hands and fingers, like a pianist would.  

An interesting complementary direction would be also to learn the correlation between music and piano keys by  training on midi files, and then combine with the method described in this paper.  

\begin{figure}
\centering
\includegraphics[width=.5\textwidth]{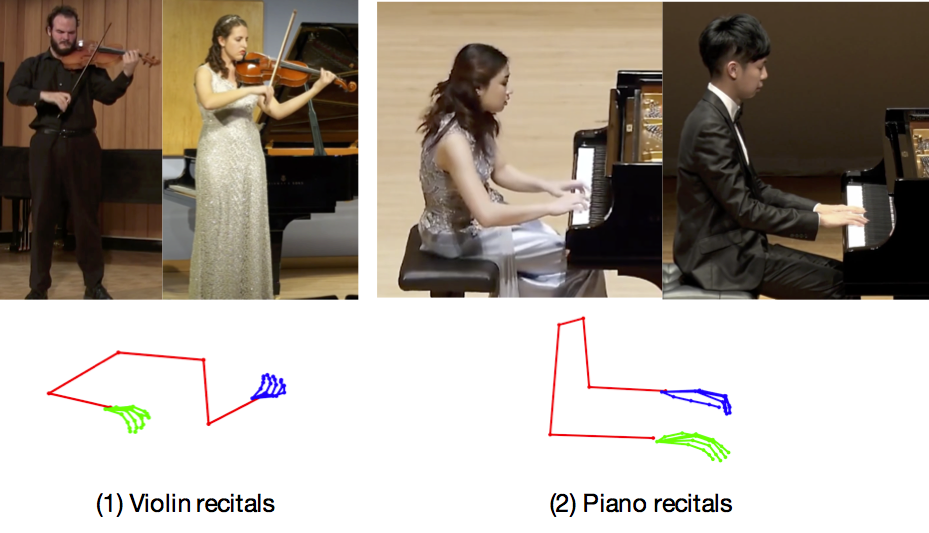}
\caption{Training data: Example frames from violin and piano sets  and their corresponding example keypoints.}
\label{fig:set_types}
\end{figure}

We consider two sets of data, piano and violin recitals (Fig.~\ref{fig:set_types}).  We collected videos for each of the two categories, and processed the videos by detecting upper body, and fingers in each frame of each video (given the points are visible).  Total of $50$ points per frame, where $21$ points represent the fingers in each hand, $8$ points for upper body. 

In addition to predicting the points, one of our  goals is to visualize the points via animation of an avatar that moves naturally according to the given audio input. We solve this problem is two steps. First, we build a Long-Short-Term-Memory (LSTM) network that learns the correlation between audio features and body skeleton landmarks. In the second part, we automatically animate an avatar using predicted landmarks. The final output is an avatar that moves according to the audio input.  

For each set we trained a separate neural network. I.e., separate network for violin and separate network for piano. The output skeletons are promising, and produce interesting body dynamics.  We encourage the reader to watch the supplementary videos with audio turned on, to experience the results.

%% file: content/related.tex
\section{Related Work}\label{sec:related}

The correlation between speech and facial movements was researched extensively beginning with classical works by \cite{brand1999voice} and \cite{bregler1997video} and most recently showing remarkable results of generating high quality videos of a face talking just from audio \cite{suwajanakorn2017synthesizing} and animating avatars using speech \cite{taylor2017deep,karras2017audio}. The three later papers have an extensive state of the art summary on facial animation from speech. 

Animating body pose just from music was not explored as far as we are aware. There is, however, a large body of work in three related areas: behavioral studies of how people's bodies react to music and sounds,  computer vision, learning and graphics research in creating and predicting natural body pose changes, for example to learn walking and dancing styles from videos (no audio), and multi-modal studies of combining audio and video input to improve recognition of facial expressions and body poses.  We will describe the related works below. 

Multi-modal studies \cite{jaimes2007multimodal} have shown that combining audio and visual inputs produces higher accuracy results than just each of the modalities alone \cite{wollmer2013lstm}. For example, facial expression recognition may benefit from also getting voice input since the emotion/pitch/loudness of the message may help recognition \cite{chen2000joint,sebe2005multimodal}.  Wang et al. showed that body pose estimation \cite{ouyang2014multi} works better with audio input \cite{wang2005inferring}. 
Multi-speaker tracking \cite{ban2017exploiting}, and combination of modalities to identify intent to interact \cite{schwarz2014combining} are additional examples. 

In dyadic communication, the relationship between speech and body rhythms was investigated, e.g.,  \cite{dittmann1972body,wagner2014gesture} with interesting results showing that certain speech characteristics were correlated with body movement frequencies \cite{boomer1964speech}. Different emotional scenarios had different types of movements and interactions, e.g., regular social interaction \cite{dittmann1969body} vs. interview \cite{boomer1964speech}. \cite{haga2008correspondences} and \cite{thompson2012exploring} researched if there is correspondence between the way pianists perform particular music, and found that there is  consistency in the way we perceive music-movement correspondences. They also demonstrate that correspondences emerge flexibly, i.e. that the same musical excerpt may correspond to different variants of movement. We find these  studies  an inspiration to our idea that body movement can be  possibly  predicted from music and speech.

There is a large body of work of the use of LSTMs to predict and edit future  poses given a short video of movement  or even a single photo, e.g.,  \cite{fragkiadaki2015recurrent,martinez2017human,habibie2017recurrent} and  \cite{holden2016deep,li2017auto,walker2017pose,chao2017forecasting}. State of the art body pose estimation and recognition techniques typically use CNNs, e.g., \cite{insafutdinov2016dense,belagiannis2017recurrent,insafutdinov2016deepercut,wei2016cpm} and \cite{he2017mask}. Earlier works showed that it is possible to learn and predict motion style \cite{liu2005learning} from videos and further animate the characters in real-time \cite{grochow2004style}.  These works did not use audio as their input. 

A complementary  area of research is  to predict audio from video. This is the inverse of our goal of  predicting video from audio. Some examples include estimating speech from face videos with the goal of lip reading \cite{chung2016lip}, or predicting which sounds an object make from video \cite{davis2014visual,owens2016visually} or photo \cite{soler2016suggesting}.  Finally, \cite{liao2015audeosynth} explored creating audio driven video montages.  

While related works do not  talk directly about animating body pose just from speech or music, it provides sufficient evidence that some correlation exists. This encourages us to explore  estimation of such correlation. We do not expect to reproduce exact body motion, and in fact assume that the transformation from speech/music to body motion is not unique but we do aim to create natural looking body pose that makes sense for the music or speech. 

%% file: content/data.tex
\section{Body Dynamics Representation}\label{sec:data}

There are variety of  classes of videos where the correlation between audio and video is interesting. E.g., videos of dance, people giving lectures, playing musical instruments, or stand-up comedy to name a few.  We chose to experiment with several of those.  Below we describe which datasets we use, how we process the audio and video signals, and how the training data is prepared and represented. The training and testing algorithm is described in Section~\ref{sec:neuralnet}.

\subsection{Data sets} 
We have experimented with data from violin recitals, and piano recitals.  The full list of URLs is available in the supplementary material and example URLs in the footnote below\footnote{Violin recital: \url{https://www.youtube.com/watch?v=BP65MiYNh50}, and Piano recital: \url{https://www.youtube.com/watch?v=SD1nhx9qYH0}}.  Figure~\ref{fig:set_types} shows example frames from each of the sets and their corresponding keypoints.  

All the videos used were downloaded from the Internet and are videos ``in the wild''. We made sure to select videos which are favorable to the processing algorithms. Specifically, the intuition behind our choice of video  was to have clear high quality music sound, no background noise, no accompanying instruments, solo performance. On the video quality side, we searched for videos of high resolution, stable fixed camera, and bright lighting.  We preferred longer videos for continuity. Our selected videos range from 3 min to 50min. We found that using recitals or single person shows are optimal sets that satisfy the above goals. Total of 3.6 hours of violin recitals was collected and 4.4 hours of piano recitals.

\subsection{Audio Processing} 

MFCC features were shown to be successful in previous art to identify and classify different musical instruments. E.g., \cite{loughran2008use} showed that  a network trained on PCA of MFCC coefficients of flute, piano and violin can recognize which instruments are used. It was also shown that MFCCs perform well for capturing the variation in speech \cite{zheng2001comparison}.  We  follow the optimal process for computing the features described in \cite{suwajanakorn2017synthesizing} with several modification to adjust to our datasets frame rate.  Specifically, we compute the features on stereo $44.1$Khz sample rate audio, perform RMS normalization to 0 db using FFMPEG  \cite{bellard2012ffmpeg}, and choose the window length as $1000/$video$_\text{fps}$ with fps$=24$, i.e., $41.66$ms. We synchronized all videos to have the same audio and video rates.  The final audio feature size is 28-D which includes the 13-D feature vector, their temporal derivative, and log mean energy for volume.

\subsection{Keypoints Estimation} 

We are interested in two types of keypoints: body, and hand fingers. Typically, estimation of keypoints on videos in the wild is challenging due to the large variation in camera, lighting, and fast movement which can divert from type of benchmark videos on which the algorithms evaluate. Recently, however a number of methods appeared that can handle in the wild videos much better. We have came up with a process that allows us to get sufficiently accurate keypoints, as described below. 

We begin by running a video through three libraries: OpenPose that provides face, body, and hands keypoints \cite{simon2017hand,wei2016cpm,cao2017realtime},  MaskRCNN \cite{he2017mask}, and DeepFace face recognition algorithm \cite{taigman2014deepface}.  Those three libraries perform well on benchmarks, but on our videos in the wild each fails on some frames.  We have noticed, however, that the failure and success are somewhat complementary across the algorithms. 

\begin{figure}
\includegraphics[width=0.5\textwidth]{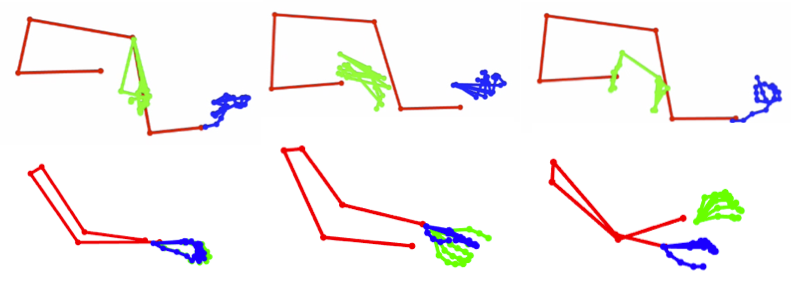}
\caption{Example failure cases of keypoints detectors (row 1 for violin set, and row 2 for piano set) which were eliminated automatically in our preprocessing step.  }\label{fig:badframes}
\end{figure}

Thus, per video we select a single frame and a detection box that includes the person of interest, as our reference frame.  The person's face in the box gets a signature from the face recognition algorithm. Each consequent   frame's box is automatically eliminated if it's 1) too far from the location of the box in the reference frame, 2) face signature doesn't match, or 3) the $L_2$ distance between points in consecutive frames is too big. See thresholds in Experiments section.  Given the box we use the hands points from OpenPose that appear in that box.  We further  choose among MaskRCNN or OpenPose which body points to use based on the confidence and existence of points (in case part of the points are not detected due to occlusions, or mis-detection). In case part of points are still missing we exclude that frame from training. See Figure~\ref{fig:badframes} for example frames that were eliminated from training.

\subsection{Keypoints Motion Factorization}
Consider a video with keypoints estimation per frame, resulting in a timeseries of points.  The motion in keypoints across frames is a product of several key components:
\begin{equation}
V_\text{frame} = M_\text{camera} ( M_{s,T,R} (V_\text{person} + V_\textrm{audio}))\label{eq:keypoints}
\end{equation}
where $M_\text{camera}$ is motion due to camera, e.g., zoom in-out and viewpoint change,  $M_{s,T,R}$ is the rigid transformation of the person, $V_\text{audio}$ is the person body transformation due to audio, e.g., bow drawing, and striking piano keys, and $V_\text{person}$ is non-rigid transformation of the person's body which is not correlated to audio, e.g., person pacing on stage while playing violin. We assume fixed stable camera, and we solve for scale, translation and rotation in 2D by choosing a reference points configuration. We assume that $V_\textrm{person}$ is zero, for simplicity. Our goal is then to predict $V_\textrm{audio}$. To reduce the dimensionality and noise of the data we compute PCA coefficients of the keypoints as follows. 

\textbf{PCA on keypoints:} Given aligned points per frame in each video, we collect all points to a single matrix of size $2p\times f$, where $p$ is the number of points per frame, each point has $2$ dimensions, and $f$ is the total number of frames in a dataset. Each set of points is reshaped to a 1-D vector of length $2p$. 
We compute PCA on the final matrix across all frames, and reduce the dimensionality to capture $90\%$ of the data. Each frame is then represented using the PCA coefficients. This allows us to reduce noise, as well as reduce dimensionality. As a final step we upsample the PCA coefficients linearly to $4\times\text{video}_\text{fps}$.  Exact  numbers per set and implementation details are also presented in Section~\ref{sec:experiments}.   

The resulted PCA coefficients (represent body motion) and MFCC coefficients (represent audio) are used in the next steps.

%% file: content/neuralnet.tex
\section{Audio to Body Keypoints Prediction}\label{sec:neuralnet}

\begin{figure}[h!]
\centering
\includegraphics[width=0.4\textwidth]{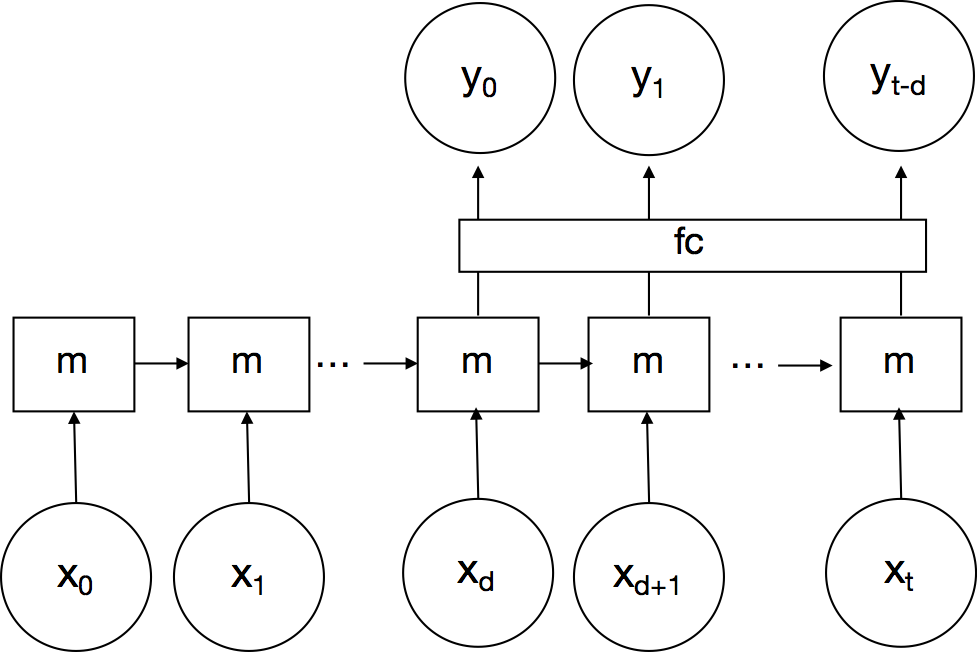}
\caption{Architecture of our points prediction LSTM. $x_i$ represent the audio features, and $y_j$ represent the corresponding keypoints.}\label{fig:lstm}
\end{figure}

Our goal is to learn a correlation between audio features and body movements. For this, we built an LSTM (Long-Short Term Memory) network \cite{hochreiter1997long}. It was shown recently that LSTMs can be used successfully to predict lip sync, e.g.,  \cite{suwajanakorn2017synthesizing} (this paper also has a nice explanation of LSTM vs. RNN vs. HMM models, and the use of LSTM with time-delay). Our architecture is visualized in Fig.~\ref{fig:lstm}. We chose to use  a unidirectional single layer LSTM with time delay. $x_i$ is an audio MFCC  in a particular time instance $i$, $y_i$s are PCA coefficients of body keypoints, and $m$ is the memory. We also add a fully connected layer 'fc' which we found to increase  performance. We have experimented with other variations of the architecture  which did not show improvement in results. Those include: using H norm for loss, global time delay, multi-layer LSTM, training on keypoints directly rather than PCA components. 

\begin{figure*}
\centering
\includegraphics[width=0.99\textwidth]{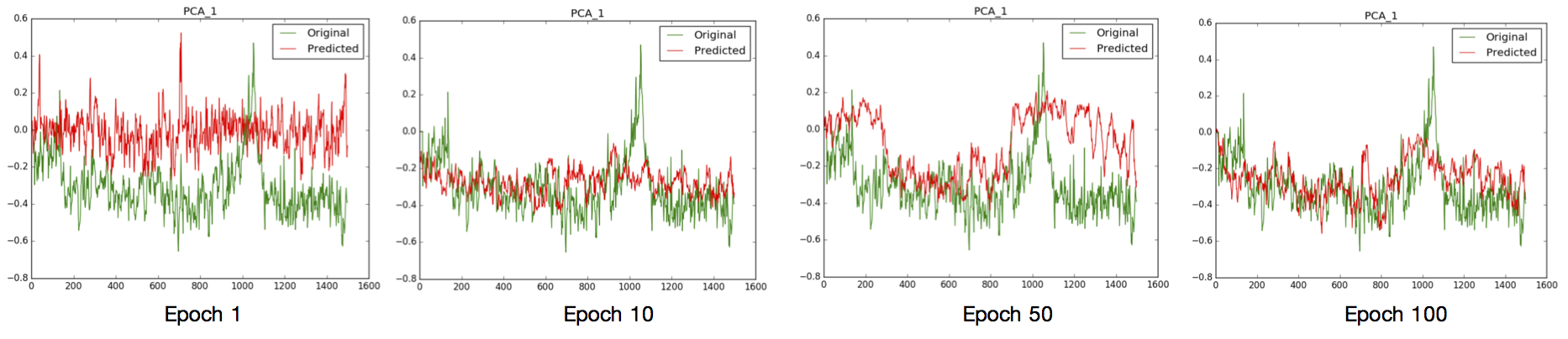}
\caption{Evolution of testing of first PCA mode (piano) as function of epochs. }\label{fig:epochs}
\end{figure*}

The parameters that we used are hidden state of $200$ (the length of $m$), trained with truncated back propagation with time steps of $400$, time delay of $5$, dropout of $0.4$, learning rate of $5e-3$. The number of PCA components we typically use is $10$. We ran the training for $300$ epochs.  The network is implemented in Caffe2 \cite{jia2014caffe}, and use ADAM optimizer. Both input and output are normalized by subtracting the mean and dividing the variance. In Fig.~\ref{fig:epochs} we show the improvement in PCA coefficients prediction as a function of epochs for the piano set. 


%% file: content/avatar.tex
\section{Body Keypoints to Avatar}\label{sec:avatar}

Once the keypoints are estimated we  animate an avatar using the points. We have built an Augmented reality application using ARkit which runs on the phone in real time.  Given a sequence of 2D predicted points, and a body avatar,  the movements are applied onto the avatar. Below we describe the specific details.

The avatars we have used are 3D body models with a body bone rig. We first initialize the rig by aligning the predicted points to 3D world coordinates. We do it by estimating average left and right shoulder points across all frames, and calculate rigid transformation to the avatar.  Then we consider each of the body, arms, head, and fingers separately. If some of the body, arms, head or fingers are not provided only the provided parts are animated. Finally, we apply root rotation offset to match pose angle of piano. The z depth for the piano is  calculated off of the wrists x position, the further left it is the closer it gets to camera. There are details below on how we rig the violin.

\textbf{Body:}  IK chain was created where the root node defined as average between the left and right hip, and connected to the average of left and right shoulder. This defines the spine. We then estimate the average spine length across the frames, and scale the avatar spine accordingly. 

\textbf{Arms:} We defined an IK chain, where the reference point was used as the wrist point. Next we calculate an offset that defines how much the arms point forward (out of fixed plane). The length of the forearm determines the offset. First, the maximum forearm length is calculated across all frames, then if the forearm appeared in its maximum length, then the reference is on the original 2D plane and arm is straight, otherwise if length is zero, then the arm is perpendicular to the source plane.


\textbf{Hands/Fingers:} 
Hand rotation was determined via root joint of the pinkie finger and the root joint of the pointer finger. E.g.,  if on the right hand the pointer joint is to stage left of the pinkie joint, then the hand must be rotated to the avatar's right with the palm facing up. The palm would be facing down if the opposite were true.  Additionally, for each finger, the angle between the reference point and root was calculated and applied. 

\textbf{Rigging the violin:} Rigging the violin was done with four points used as constraint references. The violins position was constrained to a point attached to the head and the rotation was determined using a lookat constraint the was attached to the left hand. The bow position is constrained to a point on the right hand and the rotation uses a lookat constraint attached to the bridge of the violin.


%% file: content/results.tex
\section{Experiments}\label{sec:experiments}

In this section, we discuss our experiments, implementation details, comparisons,  and evaluations.

\textbf{Running times and hardware:} Our preprocessing, train and test runs are done on a GPU server of $8$ NVIDIA M$40$ GPUs, two  $12$ cores CPUs, $256$GB RAM.   The running time for $200$ epochs of training of $3.6$ hours of violin set took $1.5$ hours (30s per epoch). Piano dataset is 4.4 hours. The running time of the test set and animation of the avatar is real time. The processing time to calculate keypoints on the training set is 10fps for OpenPose and 2fps for MaskRCNN.  

\textbf{Data:} We have experimented with Violin recitals and Piano recitals. All videos were synchronized to 24fps. We have randomly selected $20\%$ of each of the datasets for validation and $80\%$ for training. Each of the  sets was trained and tested completely independently, i.e., we have a violin net, and a piano net.  We have removed approximately $10\%$ of the frames in the videos due to not accurately predicted keypoints (based on the procedure described in Sec.~\ref{sec:data}), the total number of frames in Violin set is about 324,000,  and 414,000 for piano recitals. The threshold for removing frames are if points are farther from previous frame more than $10\%$ of the width of frame (typically 20 pixels). Test audio was not part of the training or validation sets.

\begin{figure}
\includegraphics[width=0.5\textwidth]{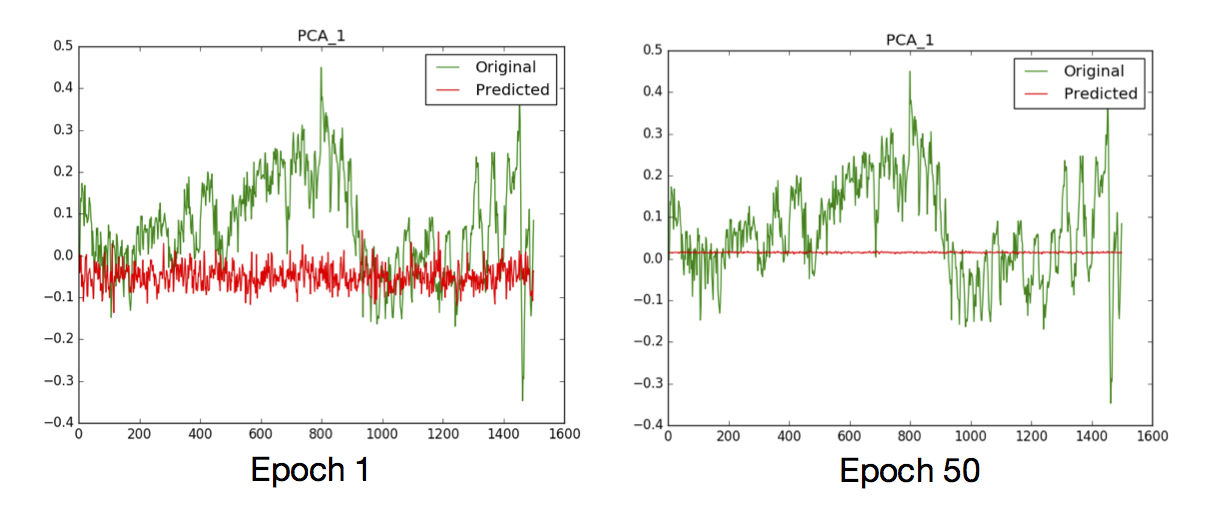}
\caption{Predicting a random sequence with violin net. We see that the prediction is stationary, means there is no correlation between the signal and violin net. } \label{fig:cross_class}
\end{figure}

\textbf{Evaluations:}  We have experimented with different parameters choices in our network and provide comparisons in Tables \ref{table:results_violin} and \ref{table:results_piano}.  To find the optimal parameters specified in Sec.~\ref{sec:neuralnet} we ran a hyper-parameters search. The errors in the tables are presented in pixels, where low is better. To achieve good results it is important to filter out all bad frames in the training data (wrong skeleton, wrong person detection, wrong person recognition), we see that errors drop significantly just by filtering out the data. In case we use half of the training data we improve the training error but test error is larger. By using less PCA coefficients we can fit better to the training data but test error is larger than using more coefficients. Using dropout doesn't improve results in our case. Time delay helps improve results.  We have experimented (Fig.~\ref{fig:cross_class}) with giving a random sequence as test audio to predict points with both nets, the test showed no correlation and no movement was predicted, as expected. 


\begin{table}
\begin{tabular}{ llll }
Method & Train & Valid &  Test\\
  \hline			
Frames not dropped & 12.44& 12.09 & 18.31\\
$50\%$ training data & 7.22 & 11.18 & 9.38\\
PCA coeff $=5$  & 6.86 & 7.04 & 9.35\\
Dropout $=0.6$  & 7.25 & 7.51& 9.31\\
Seq. length $=500$ & 7.40 & 6.95& 9.24\\
No time delay  & 7.32& 7.76& 9.19\\
Upsample 6x & 7.07 & 6.95& 9.18\\
PCA coeff $=30$  & 7.39 & 7.12 & 9.13\\
Seq. length $=200$ & 6.88 & 6.75& 9.13\\
Dropout $=0$ & 7.32	&7.71	&9.03\\
Upsample 2x & 7.33 & 9.09& 8.66\\
\textbf{Final ours} & \textbf{7.35} &  \textbf{6.58}  &  \textbf{8.31} \\
\end{tabular}
\caption{Violin net: Comparison of errors in training, validation, and testing  with different parameter choices. Errors presented in pixels (low is better). Ours is: PCA coeff = 15, Hidden variables = 200, Seq. length = 60, Batch size = 100, Learning rate 2-e03, time delay=24ms, dropout = 0.15, upsample = 4x. }\label{table:results_violin}
\end{table}

\begin{table}
\begin{tabular}{ llll }
Method & Train & Valid &  Test\\
  \hline			
Frames not dropped & 17.15  & 14.64 & 9.49 \\
$50\%$ training data  & 6.94&	6.65 &	8.87 \\
Interpolated missing data & 7.05&	6.41	&8.14\\
Upsample 2x & 7.18	& 6.95&	7.98 \\
No time delay& 6.83	& 6.04&	7.48\\
PCA coeff $=30$  & 6.66	& 6.79&	7.42\\
Dropout $=0.6$ & 6.76	&5.97&	7.38\\
Dropout $=0$ & 6.77	&6.07	&7.28\\
Seq. length $=200$ & 6.76&	6.92&	7.06\\
PCA coeff $=5$ & 6.27	&7.24	&6.94\\
\textbf{Final ours} & \textbf{6.52}	&\textbf{6.06}	&\textbf{6.84}\\
\end{tabular}
\caption{Piano net: Comparison of errors in training, validation, and testing  with different parameter choices. Errors presented in pixels (low is better).  Ours is: PCA coeff = 15, Hidden variables = 200, Seq. length = 60, Batch size = 100, Learning rate 1-e03, time delay=24ms, dropout = 0.1, upsample = 4x. }\label{table:results_piano}
\end{table}

\begin{figure*}
\includegraphics[width=0.99\textwidth]{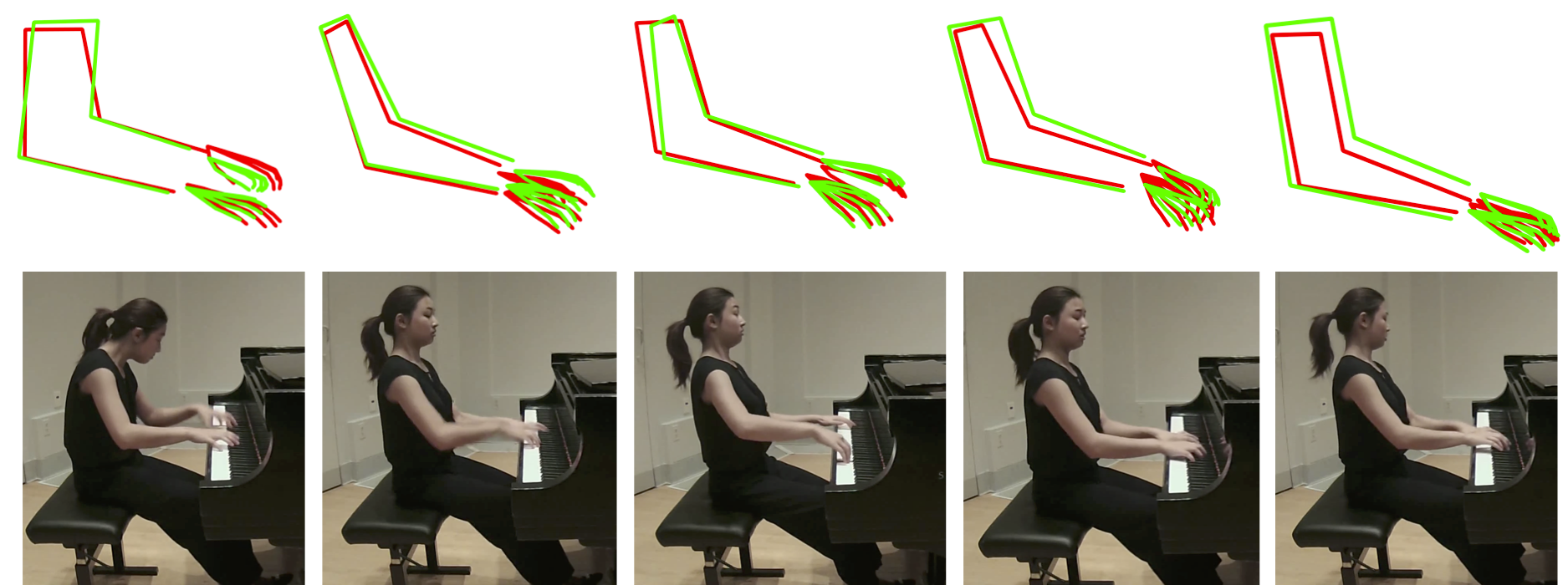}
\caption{Piano test results. We encourage the reader to watch the supplementary videos (with audio on) too. In row 1 we show predicted points from audio (in green) overlaid on top of ground truth points (red; 2d pose detector). Row 2 shows the corresponding frame for context. Note that we don't expect for them to fit exactly but just aiming for similar hands and fingers configurations. }\label{fig:piano_results}
\end{figure*}

\begin{figure*}
\includegraphics[width=0.99\textwidth]{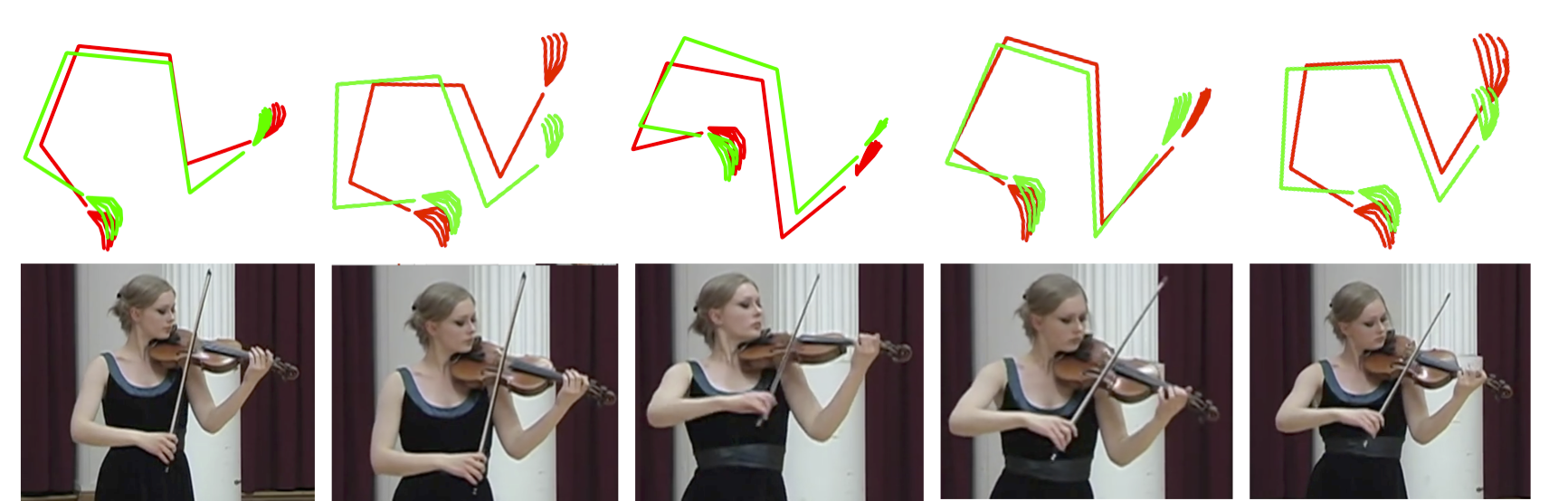}
\caption{Violin test results. We encourage the reader to watch the supplementary videos (with audio on) too. In row 1 we show predicted points from audio (in green) overlaid on top of ground truth points (red; 2d pose detector). Row 2 shows the corresponding frame for context.  }\label{fig:violin_results}
\end{figure*}

\textbf{Results:} In Figure~\ref{fig:piano_results} and \ref{fig:violin_results} we present representative results. We show the predicted keypoints for different body poses,  as well as the original frame for context. For the keypoints we show them overlaid on  the ground truth points for visual comparison. Note that we don't expect the points to be exactly the same, but the fingers and hands to produce a similar satisfactory movement which was the goal of this paper. The ground truth in our case is result of 2D body pose detector which can  be mistaken.   Finally, we show failure cases in Fig ~\ref{fig:failure}, row 1 shows piano, and row 2 violin.  These show limitations of our system: currently our system is trained on 2D poses, while actual poses in training videos are 3D. Consequently, occlusions and invisible points are not predicted well. In high pace and frequency parts of the videos, the body pose detectors may create mistakes, similarly in case of motion blur. This causes the network to learn behave in high frequency audio accordingly. 

\begin{figure*}
\includegraphics[width=0.99\textwidth]{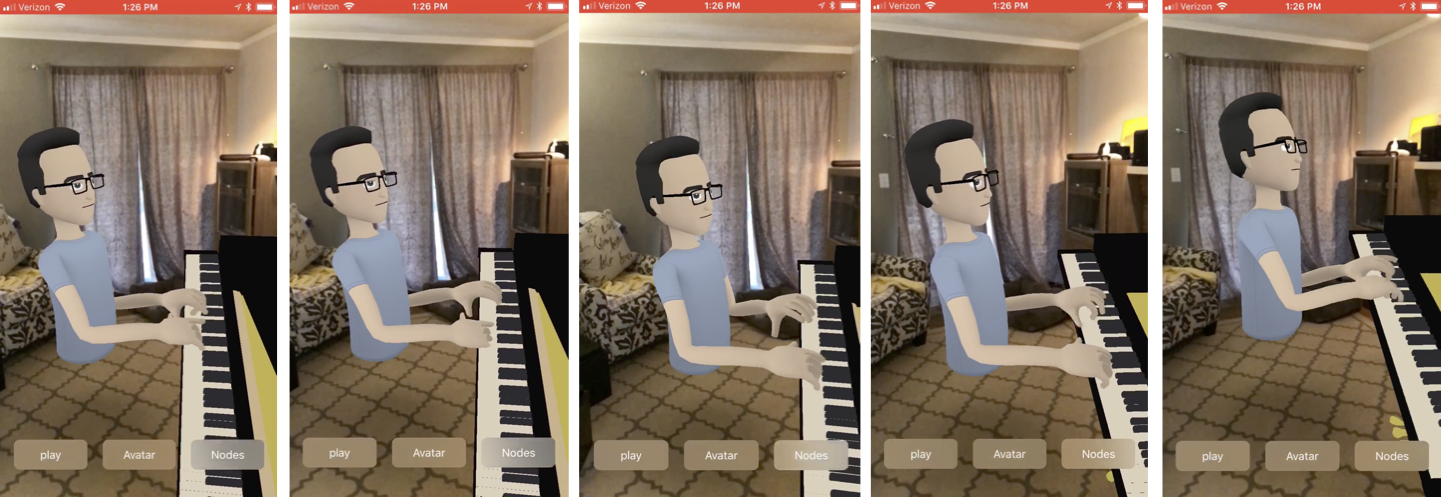}
\caption{Avatar playing piano: Screen shots of avatar animation result with the predicted skeleton.  }\label{fig:piano_avatar}
\end{figure*}

\begin{figure*}
\includegraphics[width=0.99\textwidth]{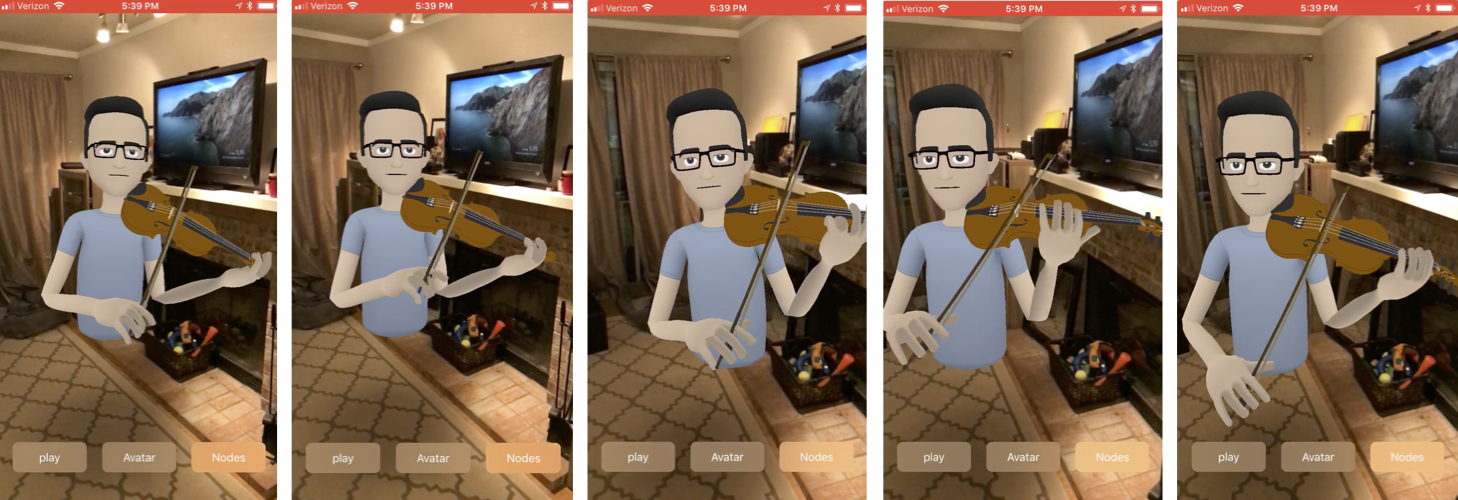}
\caption{Avatar playing violin: Screen shots of avatar animation result with the predicted skeleton.  }\label{fig:violin_avatar}
\end{figure*}

Figures \ref{fig:piano_avatar}  and \ref{fig:violin_avatar} show screenshots of our avatar animation based on predicted points. The piano, violin, and avatar are synthetic objects placed in a real scene (augmented with ARKit).

\textbf{Videos:} We recommend watching the supplementary video with audio on. In addition to showing the results we present how an animation may look like when music and keypoints don't match, you will see in the video that it's look very unnatural thus the importance of the network.

%% file: content/conclusion.tex
\section{Discussion and Limitations}

We have proposed a new hypothesis that body gestures can be  predicted from audio signal, and  showed promising initial results.  We believe the correlation between audio to human body  is  very promising for a variety of applications in VR/AR and recognition. It was shown previously that mouth animation can be done just from audio, and in this paper we show initial results on body animation. We hope it will open up  further  research. There are a number of limitations and many extensions that would be interesting to explore. 

One direction is to enable 3D movement, currently we use OpenPose and MaskRCNN to estimate keypoints. These are both 2D points estimators.   Provided a 3D body keypoints estimator, e.g., Vnect \cite{mehta2017vnect} or SMPL \cite{bogo2016keep} our approach can potentially be extended to handle 3D motion as well, and allow more diverse human modeling. 

Another direction is to predict occluded keypoints. Currently, we can only predict the visible points and if training frames do not have all the points we ignore that frame. It would be interesting to see if the keypoints from audio network can be extended to predict occlusions, e.g., in piano the left  hand usually occludes the right hand but the audio includes music played by both hands. 

We have used only YouTube videos as training data. In the future, to increase realism and accuracy we could consider complementing the training data with sensor information or midi files. 

Finally, getting good training data per class of activity is not straight forward. Note the constraints we required in Sec.~\ref{sec:data}.  It would be interesting to explore a general network that can handle variety of poses, without the need to classify the type of action a priori. Alternatively, it would be interesting to incorporate video activity recognition into our learning framework.

\begin{figure}
\includegraphics[width=0.5\textwidth]{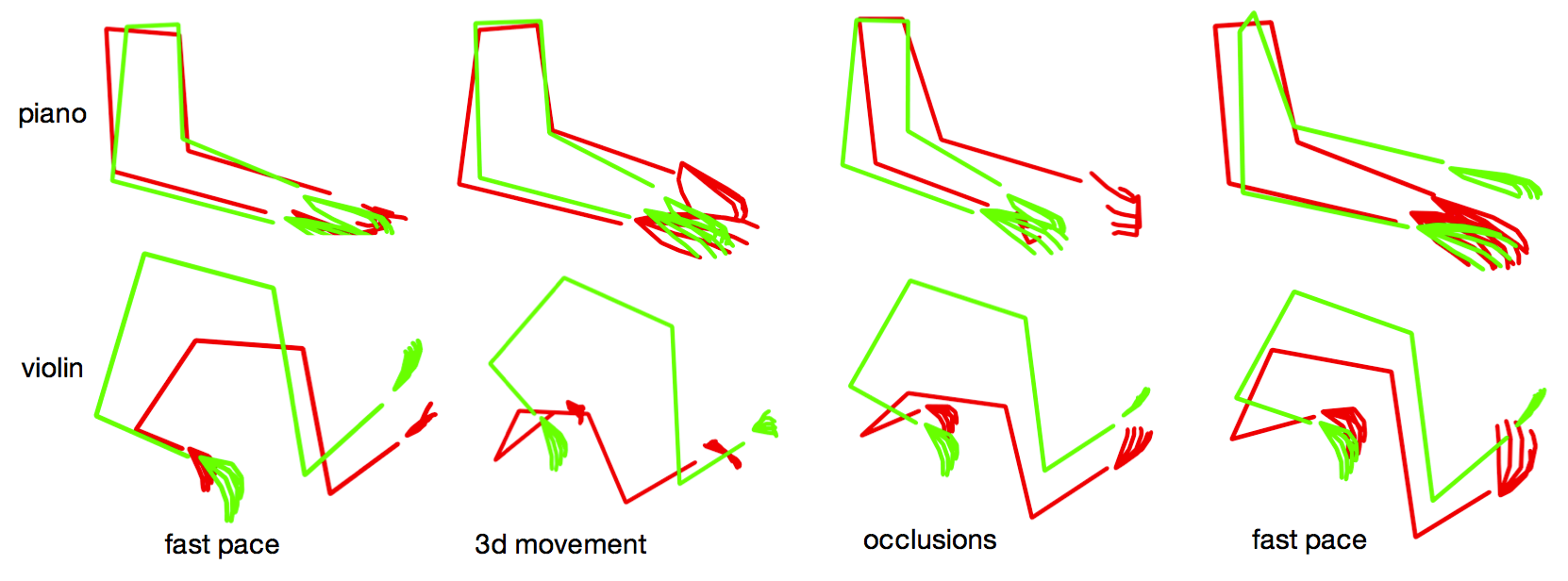}
\caption{Typical limitations of the method. The method fails with extreme occlusions, or unexpected tones or movements that are not captured well in the training data (fast raising of arms, movement from one side of keyboard to extreme other, or very fast pace). It may also fail since we train based on 2D pose estimators while people in  YouTube videos move in 3D. The predicted points (green) are overlaid on top of ground truth (red) for comparison, in those cases also the ground truth points often (estimated by pose detection) are not exact. }\label{fig:failure}
\end{figure}